\documentclass[aps,prl,epsfig,floats,twocolumn,superscriptaddress,amssymb,amsmath,floatfix,showpacs]{revtex4}
\usepackage{chemarr}
\usepackage{graphicx}
\usepackage{bm}  
\usepackage{amssymb}
\usepackage{color}
\usepackage{amscd}

\begin{document}

\title{Mechanical Response of a Small Swimmer Driven by Conformational Transitions}

\author{Ramin Golestanian}
\email{r.golestanian@sheffield.ac.uk} \affiliation{Department of
Physics and Astronomy, University of Sheffield, Sheffield S3 7RH,
UK}

\author{Armand Ajdari}
\affiliation{Gulliver, UMR CNRS 7083, ESPCI, 10 rue Vauquelin, 75005
Paris, France}

\date{\today}

\begin{abstract}
A conformation space kinetic model is constructed to drive the
deformation cycle of a three-sphere swimmer to achieve propulsion at
low Reynolds number. We analyze the effect of an external load on
the performance of this kinetic swimmer, and show that it depends
sensitively on where the force is exerted, so that there is no
general force--velocity relation. We discuss how the conformational
cycle of such swimmers should be designed to increase their
performance in resisting forces applied at specific points.
\end{abstract}

\maketitle



Active transport is a most fascinating aspect of the busy life in
the cell \cite{howard}. In the nano-scale world where thermal
agitations are wild, miniature machines called molecular motors
convert chemical energy---from hydrolysis of ATP
molecules---directly into useful mechanical work, in the form of
carrying cargo or sliding actin filaments along one another. While
it is difficult to imagine fabricating such sophisticated machines
in the lab, one may naturally wonder if it is possible to design
simpler machines with similar functionalities \cite{leigh-etal}. In
this flavor, an interesting target is an autonomous small scale
swimmer \cite{Dreyfus,phoretic}, which could later on be steered by
coupling to a guiding network or system.

Swimmers at small scale (low Reynolds number) have to undergo
non-reciprocal deformations to break the time-reversal symmetry and
achieve propulsion \cite{taylor}. This imposes significant
constraints when one wants to design a swimmer with only a few
degrees of freedom and strike a balance between simplicity and
functionality \cite{purcell1}. Recently, there has been an increased
interest in such designs \cite{3SS,swimmer} and an interesting
example of such robotic micro-swimmers has been realized
experimentally using magnetic colloids attached by DNA-linkers
\cite{Dreyfus}.

Here we combine features of simple low Reynolds number hydrodynamic
swimmers and elements characteristic of models for chemical
molecular motors. We focus on a recently introduced three-sphere
swimmer \cite{3SS} with the minimal two degrees of freedom. Instead
of assuming a prescribed sequence of deformations, we consider these
deformations to occur stochastically, as conformational transitions
between elongated and shortened {\em states} for each of the two
degrees of freedom. This gives us a swimmer with a velocity that
depends on the transition rates between these states, which in
practice could come about via mechanochemical transitions, i.e. due
to chemical reactions that are coupled with such mechanical
deformations \cite{howard}. We check that a net velocity requires
that detailed balance in the transition rates is broken. Using this
simple kinetic model, we study the effect on the swimming velocity
of a resisting external force or load: the load clearly drags the
swimmer backwards, {\em but also} puts elements in compression or
extension, thereby modifying the transition rates between extended
and shortened states. As a consequence, we find that the performance
of the motor strongly depends on where the force is exerted, in
contrast to the usual perception that the performance of a swimmer -
or a motor - can be summarized in a unique force--velocity relation.
Interestingly, the motor performance can in some special cases be
increased upon application of the external load, provided it is
applied at the right location. More generally, we discuss efficient
strategies for optimizing the performance of this swimmer.

\begin{figure}[b]
\includegraphics[width=.8\columnwidth]{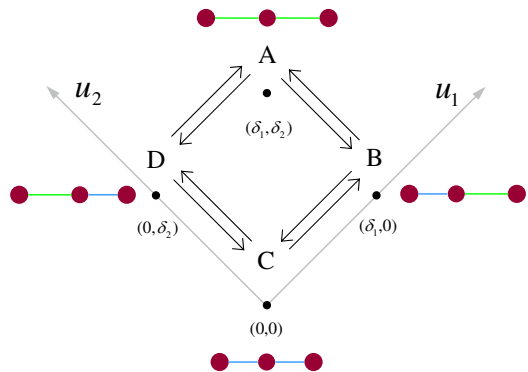}
\caption{Schematics of the configuration space of the three-sphere
swimmer. To maintain a net swimming to the right, the deformation
moves need to make more clockwise full cycles than counterclockwise
ones.} \label{fig:config}
\end{figure}

\begin{figure}[t]
\includegraphics[width=.8\columnwidth]{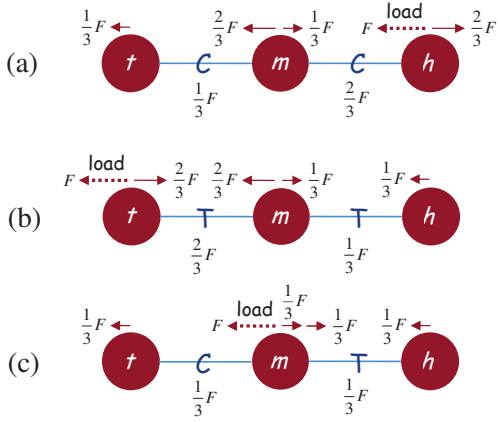}
\caption{Force balance on the spheres and the linkers for a swimmer
moving to the right, when the external load $F$ (acting to the
right) is attached to the (a) {\em head}, (b) {\em tail}, or (c)
{\em middle} of the swimmer. In each case, it is identified whether
each linker is under {\em compression} (C) or {\em tension} (T),
with the value of the force given (underneath).} \label{fig:load}
\end{figure}

We start with the swimmer model introduced in \cite{3SS}, namely
three spheres connected by two linkers of negligible hydrodynamic
effect, that cycle in time between extended and short states. For
simplicity, we take here all sphere radii to be equal to $a$. We
further assume that the lengths of the two arms are $L_1(t)=\ell_1 +
u_1(t)$ and $L_2(t)=\ell_2+u_2(t)$ with the $u_i$'s being small
perturbations about the average lengths. For prescribed arm
deformations, writing a force balance on each sphere leads an
instantaneous net displacement velocity of the swimmer, that can
here  be written as a series expansion
\begin{math}
v(t)=A_i {\dot u}_i+B_{ij} {\dot u}_i {u}_j+C_{ijk} {\dot u}_i {u}_j
{u}_k+\cdots,
\end{math}
where the coefficients $A_i$, $B_{ij}$, $C_{ijk}$, {\em etc.} are
purely geometrical prefactors ({\em i.e.} involving only the length
scales $a$ and $\ell_i$). After many cycles, this process gives a
vanishing contribution from the linear terms ${\dot u}_1$ and ${\dot
u}_2$ and from the symmetric combination ${\dot u}_1 u_2+{\dot u}_2
u_1=d (u_1 u_2)/dt$. Thus to leading order the average swimming
velocity is
\begin{equation}
V \equiv \langle v \rangle=\frac{K}{2} \langle{\dot u}_1 u_2 - {\dot
u}_2 u_1\rangle=K \left \langle \frac{d {\cal A}}{d t} \right
\rangle,\label{eq:v-def}
\end{equation}
where $d {\cal A}$ is the area element in the $(u_1,u_2)$ space, and
\begin{math} K=\frac{a}{3}
\left[\frac{1}{\ell_1^2}+\frac{1}{\ell_2^2}-\frac{1}{(\ell_1+\ell_2)^2}\right]
\end{math}
\cite{ag}. In other words, to the leading order the swimming
velocity is proportional to the area enclosed by the orbit of the
cyclic motion in the configuration space of the deformations.

We now focus on a situation where the two arms can be in two states
with deformations of either $u_i=0$ or $u_i=\delta_i$, and transit
from one to the other in an almost instantaneous fashion. This means
that the configuration space of the swimmer will be made of only
four distinct states as shown in Fig. \ref{fig:config}, which
correspond to different values of the pair $(u_1,u_2)$, namely:
state A for $(\delta_1,\delta_2)$, state B for $(\delta_1,0)$, state
C for $(0,0)$, and state D for $(0,\delta_2)$. We then assign
transition rates to the system, corresponding to the average rate of
opening and closing of the arms. For example, the transition rate
from state A to state B is denoted as $k_{BA}$, and similar
notations are used for the 8 rates describing forward and reverse
transitions along the cycle
\begin{equation}
{\rm A}\xrightleftharpoons[k_{AB}]{k_{BA}} {\rm
B}\xrightleftharpoons[k_{BC}]{k_{CB}}{\rm
C}\xrightleftharpoons[k_{CD}]{k_{DC}}{\rm
D}\xrightleftharpoons[k_{DA}]{k_{AD}}{\rm A}.
\end{equation}
For simplicity, and at the cost of motor efficiency, we assume that
the transitions occur quite rapidly and seldom, so that they never
``overlap.''

We can now calculate the swimming velocity as a function the
transition rates. At steady state, the average swimming velocity of
the object is given by the probability current $J$ along the
A$\rightarrow$B$\rightarrow$C$\rightarrow$D$\rightarrow$A cycle
times the net displacement while performing the cycle. This distance
$\Delta x$ is simply $K\delta_1\delta_2$, which yields
\begin{math}
V= K\delta_1\delta_2 J.
\end{math}
The probability current $J$ is a function of the transition rates,
which can be obtained from straightforward algebra:

\begin{figure*}
\includegraphics[width=2.05\columnwidth]{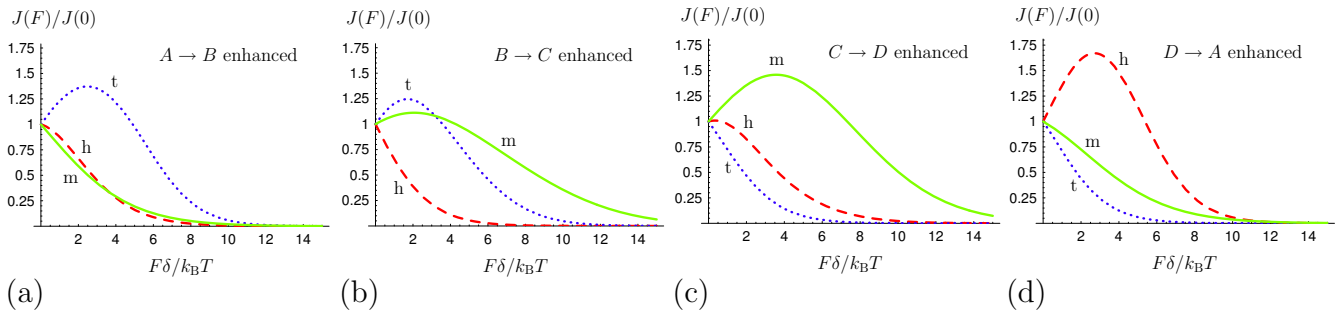}
\caption{Motor performance function $J(F)/J(0)$ versus the external
force, when only one of the rates is enhanced to $1+\epsilon$ while
the rest are kept fixed at $1$. The plots correspond to the enhanced
rate being (a) $k_{0BA}$, (b) $k_{0CB}$, (c) $k_{0DC}$, and (d)
$k_{0AD}$, and $\epsilon=50$. In each case, the dashed (red) line
corresponds to the load being attached to the {\em head}, the dotted
(blue) line corresponds to attachment to the {\em tail}, and the
solid (green) line is for the {\em middle} attachment.}
\label{fig:v-f}
\end{figure*}


\begin{widetext}
\begin{equation}
J=\frac {k_{AD}k_{DC}k_{CB}k_{BA}-k_{AB}k_{BC}k_{CD}k_{DA} } {\sum
_{\rm replace \;A \;by \;B,\;C,\;D} (k_{AD}k_{DC}k_{CB}
+k_{AB}k_{BC}k_{CD} +
k_{AB}k_{AD}k_{DC}+k_{AD}k_{AB}k_{BC})}.\label{eq:J}
\end{equation}
\end{widetext}
From the above equation it is clear that if detailed balance holds,
then $J$ is zero as the numerator vanishes. Using the average steady
state current, we can deduce the average period of one full cyclic
motion along
A$\rightarrow$B$\rightarrow$C$\rightarrow$D$\rightarrow$A as
$T=J^{-1}$. In general a $1\longleftrightarrow2$ asymmetry in the
system, together with breaking the detailed balance at least for one
of the transitions, will lead to net motion. For the particular
limit where the forward rates are all much higher than the
corresponding backward ones ($k_{BA} \gg k_{AB}$, {\em etc.}), we
find
\begin{math}
T=k_{AD}^{-1}+k_{DC}^{-1}+k_{CB}^{-1}+k_{BA}^{-1},
\end{math}
which simply means that the period for a full cycle is the sum of
the time intervals needed to complete each leg of the cycle. As
another example, we can assume that all of the equilibrium $k_{\beta
\alpha}$'s are equal to $1$ (for the sake of illustration), and that
by external action only one of them is modified {\em e.g.}
$k_{BA}=1+\epsilon$. In this case, it is easy to show that Eq.
(\ref{eq:J}) yields $J=\epsilon/(16+6\epsilon)$, which leads to a
velocity proportional to the perturbation if the latter is small and
independent of it if the perturbation is very large, as the cycling
is then limited by the other three unperturbed transitions. In
general, it is easy to see that the slowest leg of the reaction
controls the average rate of full cyclic motion.

It is interesting to study the effect of an external load on the
velocity of the system and the performance of the motor. When the
swimmer is subject to external forces, because of carrying a cargo
for example, there are two types of mechanical responses in the
system. First, the external forces enter the hydrodynamic force
balance on each sphere, and this will introduce a Stokes drag on the
sphere as a whole, which is a linearly decaying contribution to the
net swimming velocity as a function of force. Second,
the transition rates that control the kinetics of the deformations
for the two arms of the swimmer are affected by the external forces
as they will have to do mechanical work against them to induce the
deformations. Depending on where the force is applied, different
legs of the kinetic cycle could be affected, and this could lead to
a complex mechanical response with the performance of the motor
depending on the location of the load. The force-dependent kinetic
rates will yield a net current $J(F)$, which combines with the
Stokes response to give the swimming velocity as
\begin{equation}
V(F)=-\frac{F}{18 \pi \eta a_R}+V_0\;J(F)/J(0),\label{eq:V-F-def}
\end{equation}
where $V_0$ is the swimming velocity at zero force, $\eta$ is the
viscosity of the solvent, and $a_R$ is a renormalized hydrodynamic
radius \cite{note}.

The transition rates are modified in the presence of external
forces, because the mechanical energy enters the balance of
probability of the different states and transitions among them. If
there is a transition from $\alpha \rightarrow \beta$ that
corresponds to an extension by a factor of $\delta$, then under a
positive tension $f$ the rate of $\alpha \rightarrow \beta$
transitions is increased by a factor of $\exp(f x/k_{\rm B}T)$ while
the rate of $\beta \rightarrow \alpha$ transition is decreased by a
factor of $\exp(-f x'/k_{\rm B}T)$ where typically $x=\theta \delta$
is the distance between state-$\alpha$ and the energy barrier and
$x'=(1-\theta) \delta$ is the distance between the energy barrier
and state-$\beta$ ($\theta$ between $0$ and $1$). Thus the ratio
between the two transitions rates ({\em i.e.} the $\alpha
\rightarrow \beta$ rate divided by the $\beta \rightarrow \alpha$
rate) changes under a tension $f$ by a factor of $\exp(f d/k_{\rm
B}T)$ as required by the Boltzmann formula (equilibrium populations
between state-$\alpha$ and state-$\beta$ under $f$).

\begin{table}[b]
\caption{\label{tab:fba} The algebraic force $f_{\beta \alpha}$
which should be used in Eq. (\ref{eq:kba}) to calculate the forward
rates. The values for the corresponding reverse rates can be
obtained via $f_{\beta \alpha}=-f_{\alpha \beta}$.}
\begin{ruledtabular}
\begin{tabular}{cccc}
transition & head & tail
& middle \\
\hline $A \rightarrow B$ & $+\frac{1}{3} F$ & $-\frac{2}{3} F$ & $+\frac{1}{3} F$ \\
$B \rightarrow C$ & $+\frac{2}{3} F$ & $-\frac{1}{3} F$ &
$-\frac{1}{3}
F$\\
$C \rightarrow D$ & $-\frac{1}{3} F$ & $+\frac{2}{3} F$ &
$-\frac{1}{3}
F$\\
$D \rightarrow A$ & $-\frac{2}{3} F$ & $+\frac{1}{3} F$ & $+\frac{1}{3} F$\\
\end{tabular}
\end{ruledtabular}
\end{table}

In our system, the value of the force under which each arm should
close or open depends on where the load is applied. Figure
\ref{fig:load} shows the break down of the mechanical force balance
on each sphere, and the corresponding forces endured by each linker,
for the three different positions of the load. When the resisting
force is attached to the head of the swimmer, both linkers are under
compressional forces, and the compression force on the right
arm---nearer to the load---is larger than that of the left arm by a
factor of two. Attaching the load at the tail creates a similar
pattern of tensional forces. If the force acts on the middle sphere,
the left arm is under compression and the right arm is under
tension.

Using the above definition, the transition rates from a conformation
state $\alpha$ to another state $\beta$ can be written as
\begin{equation}
k_{\beta \alpha}=k_{0 \beta \alpha} \;\exp
\left(\frac{1}{2}\frac{f_{\beta \alpha} \delta_i}{k_{\rm
B}T}\right),\label{eq:kba}
\end{equation}
where $f_{\beta \alpha}$ is the force endured by the linker $i$ that
undergoes a deformation during the $\alpha \rightarrow \beta$
transition, and $\theta=1/2$ is assumed for simplicity. The sign of
$f_{\beta \alpha}$ is determined by whether the transition
(deformation) is helped $(+)$ or opposed $(-)$ by the force acting
on the linker. The values of $f_{\beta \alpha}$ are given in Table
\ref{tab:fba} for the forward reaction rates for the different
locations of the load. Note that by definition $f_{\beta
\alpha}=-f_{\alpha \beta}$, which can be readily used to calculate
the reverse rates.

The force-dependent rates [from Eq. (\ref{eq:kba}) and Table
\ref{tab:fba}] can be used in Eq. (\ref{eq:J}) to calculate the
current $J(F)$, which determines the swimming velocity under the
effect of an external load $F$. From Eq. (\ref{eq:V-F-def}), it
appears that the normalized current $J(F)/J(0)$ is a quantitative
measure of how the ability of the motor to generate propulsion is
affected by the presence of the load. In Fig. \ref{fig:v-f}, this
``motor performance function'' is plotted against the external force
for the particular example discussed above (in which only one of the
forward rates is enhanced to $1+\epsilon$ while the rest of the
rates are set to unity), and $\delta_1=\delta_2 \equiv \delta$ is
assumed for simplicity. Figure \ref{fig:v-f}a corresponds to when
the $A \rightarrow B$ (contraction of the left arm) transition rate
is enhanced, and it shows that attaching the load to the head or the
middle for both of which the left arm is under a compression of
$\frac{1}{3} F$ quickly decreases the performance of motor. On the
other hand, attaching the load to the tail of the swimmer, which
puts the left arm under a tension of $\frac{2}{3} F$, actually helps
the motor initially for forces of up to $3 k_{\rm B}T/\delta$ or so,
before eventually hampering the performance at large forces. One
notes that the force across the left arm actually helps the $A
\rightarrow B$ transition when the load is at the head or the
middle, and opposes it when it is at the tail. It thus seems that
the performance of the motor is best when the rate is enhanced for
the deformation which is most hampered by the external load. In
other words, the best strategy seems to be to try and make the
performances of the different legs of the reaction cycle as uniform
as possible, as the total velocity is controlled by the weakest
performance in the cycle. The same pattern can be seen in Figs.
\ref{fig:v-f}b--\ref{fig:v-f}d. Another interesting feature that can
be seen is that when the load is at the middle and the condition is
right for improved performance (see above) the system seems to
endure comparatively much stronger forces: in Figs. \ref{fig:v-f}b
and \ref{fig:v-f}c one can see that the performance is significant
for loads of up to about $12 k_{\rm B}T/\delta$. This is presumably
because attaching the load to the middle creates a more balanced
distribution of the forces in the linkers (still of opposite nature
but of equal magnitudes; see Fig. \ref{fig:load} and Table
\ref{tab:fba}).

Even when the performance of the motor is increased by the opposing
force, one still has a decreasing trend for the swimming velocity
because of the Stokes drag term in Eq. (\ref{eq:V-F-def}). Using a
linear approximation for $J(F) \simeq J_0 (1+c F \delta/k_{\rm B}T)$
at small forces (where $c$ is a positive constant of order unity),
one can write Eq. (\ref{eq:V-F-def}) as
\begin{math}
V(F)=V_0 \left[1-\left(\frac{1}{18 \pi \eta a_R V_0}-\frac{c
\delta}{k_{\rm B}T}\right) F\right],
\end{math}
which implies that for forces much smaller than the thermal
activation force $k_{\rm B}T/\delta$ the increased motor performance
can lead to increased swimming velocity if the viscous drag on the
swimmer is larger than the thermal activation force. While this
could be extremely difficult to achieve as it requires
unrealistically high swimming velocities, it is an interesting
fundamental possibility that increased swimming velocity can be
achieved upon exerting opposing forces.

In conclusion, we have proposed and studied a simple model of a low
Reynolds number swimmer driven by a kinetic engine. The main result
is that the ability of this swimmer to carry a load or to resist an
opposing force depends on where the load or the force is applied.
This is not linked to the stochastic nature of the present motor,
but also holds for a motor driven by a prescribed sequence of
internal stresses (to which the applied stresses add up).
Altogether, this shows that the description of such machines can go
beyond a simple force--velocity relation, more complex and maybe
richer in functionality.



\begin{thebibliography}{99}
\bibitem{note}
For the renormalized hydrodynamic radius in Eq. (\ref{eq:V-F-def}),
we find
\begin{math}
\frac{1}{a_R}=\frac{1}{a}+\langle\frac{1}{L_1}\rangle+\langle\frac{1}{L_2}\rangle
+\langle\frac{1}{L_1+L_2}\rangle-\frac{a}{2}\langle\left(\frac{1}{L_1}-\frac{1}{L_2}\right)^2\rangle
-\frac{a}{2}\langle\frac{1}{(L_1+L_2)^2}\rangle,
\end{math}
which should be expanded in 2nd order of the deformations to keep
the same order of terms in the perturbation theory.
\end{thebibliography}
\end{document}